\documentstyle[12pt,aps,eqsecnum]{revtex}
\begin{document} 
\draft
\date{\today} 
\newcommand{\BEQ}{\begin{equation}}    % Gleichungen Anfang ..
\newcommand{\BEA}{\begin{eqnarray}}
\newcommand{\EEQ}{\end{equation}}      % .. und Ende
\newcommand{\EEA}{\end{eqnarray}}
\newcommand{\eps}{\epsilon}                      % epsilon
\newcommand{\lmb}{\lambda}                       % lambda
\newcommand{\sig}{\sigma}                        % sigma
\newcommand{\vph}{\varphi}                       % rundes phi
\newcommand{\ups}{\Upsilon}                      % Gr. grosses Ypsilon
\newcommand{\rar}{\rightarrow}                   % Pfeil nach rechts
\newcommand{\tia}[1]{\tilde{a}_{#1}}             % a Tilde mit Platz fuer Index
\newcommand{\ket}[1]{\left|#1\right\rangle}      % Ket-Zustand
\newcommand{\bra}[1]{\left\langle #1\right|}     % Bra-Zustand
\newcommand{\rcd}[1]{\bar{\cal D}_{#1}}          % Renyi Kodimension D_q
                                                 % mit q als Parameter
\newcommand{\vm}[1]{\check{#1}}                  % Hut verkehrt fuer Math.
\newcommand{\zeile}[1]{\vskip #1 \baselineskip}  % N Zeilen ueberschlagen
                                                 % mit \zeile{N}
\newcommand{\vekz}[2]
     {\mbox{${\begin{array}{c} #1  \\ #2 \end{array}}$}}
                                  % \vekz{a}{b} erzeugt einen zweikomponentigen
                                  % Vektor mit den Elementen a,b.
\newcommand{\matz}[4]
     {\mbox{${\begin{array}{cc} #1 & #2  \\ #3 & #4 \end{array}}$}}
                                  % \matz{a}{b}{c}{d} erzeugt eine 2*2
                                  % Matrix mit den Elementen a,b;c,d.
 
\newcommand{\build}[3]{\mathrel{\mathop{\kern 0pt#1}\limits_{#2}^{#3} }}
                                  % Chem. Reaktionen
 
\newcommand{\appsection}{\setcounter{equation}{0} \section*{Appendix}
\renewcommand{\theequation}{A.\arabic{equation}}
              \renewcommand{\thesection}{A} }
                                    % Anhang
%\def\thefootnote{\fnsymbol{footnote}} % Aenderung der Fussnotensymbole
                        %
 
\catcode`\@=11
\def\numberbysection{\@addtoreset{equation}{section}
        \def\theequation{\thesection.\arabic{equation}}}
                        % Nummerierung pro section
%\numberbysection

\baselineskip 0.3in
\title{On the phase diagram of branched polymer collapse}

\author{Malte Henkel$^{a,b}$ and Flavio Seno$^{c}$}

\address{
($^{a}$)  Theoretical Physics, Department of Physics, 
University of Oxford, \\1 Keble Road, Oxford OX1 3NP, England} 

\address{
($^{b}$)  Laboratoire de Physique du Solide\footnote{Unit\'e de Recherche
associ\'ee au CRNS no 155}, Universit\'e Henri Poincar\'e Nancy I, BP 239, \\
F - 54506~Vand{\oe}uvre l\`es Nancy Cedex, France\footnote{permanent adress}
}  

\address{
($^{c}$)  INFM-Dipartimento di Fisica, Universit\`a di Padova, \\
Via Marzolo 8, I - 35100 Padova, Italy 
}

\maketitle

\begin{abstract}

The phase diagram of the collapse of a two-dimensional infinite 
branched polymer interacting
with the solvent and with itself through contact interactions is studied
from the $q\to 1$ limit of an extension of the $q-$ states Potts model.
Exact solution on the Bethe lattice and 
Migdal-Kadanoff renormalization group calculations
show that there is a line of $\theta$ transitions from the extended to
a single compact phase. The $\theta$ line, governed by three different
fixed points, consists of two lines of extended--compact transitions which
are in different universality classes and meet in a multicritical point.

On the other hand, directed branched polymers are shown to be completely
determined by the strongly embedded case and there is a single $\theta$ 
transition which is in the directed percolation universality class.   
\end{abstract}
\vskip 1truecm
\noindent
PACS:05.70.Jk, 64.60.Ak, 36.20.Ey, 64.60.Fr 

\newpage

\section{Introduction}

A long-standing problem in the theory of polymer molecules in solution
is the effect of intramolecular forces on the shape and size
of an isolated polymer. The intramolecular 
forces are usually assumed \cite{Flory} to 
be of van der Waals' type, consisting
of strong, short-range repulsive and weak, 
long-range attractive interactions.
Depending on temperature or solvent composition, the macromolecule
forms either extended structures or collapses into a dense globule
\cite{deGennes79}. The transition between these two states,
which is called $\theta$
transition, has attracted a great deal of attention, both
theoretically \cite{deGe75,Priv86,Sale86,Coni86,Seno88} 
\cite{Madr90,Fles92,Van93,Seno94,Stel94,Fles94} 
and experimentally partly 
because of its connection with protein folding \cite{Moore77,Volk77,Obu86}.

For the bidimensional {\em linear} polymers many theoretical tools,
such as conformal invariance \cite{Cardy90,Christe93} and Coulomb gas 
\cite{Nienhuis87} methods, can be successfully
employed and considerable progress has been 
made in understanding the nature and the critical exponents at the collapse
transition \cite{DS87,VSS91}.

In contrast much less is known about {\em branched} polymers. In
lattice statistical mechanics they  are described by
lattice animals, graphs of connected occupied sites on a  lattice.
Some examples are shown in Fig. 1. {\em Contact} interactions are introduced
between nearest neighbour sites which are not immediately linked
by a bond and {\em solvent} interactions are introduced between an
occupied site and any nearest neighbour empty sites. In 
$d$ dimensions lattice animal exponents are related to
the exponents of the Lee-Yang edge singularity of the Ising model in $d-2$
dimensions \cite{PS81}. However, this elegant theoretical approach can 
give only the bulk
entropic exponent for the extended phase. Moreover, branched polymers in two
dimensions are in general not conformally invariant \cite{Miller93}.
This model can be reformulated as the $q \to 1$ limit of an appropriate
$q$--state Potts model \cite{Giri77,Wu82}. However  this formulation has
not given any exact results \cite{Coni83} but only allowed
the conjecture that lattice animals could present two
distinct branches of $\theta$ behaviour, separated by a percolation
critical point \cite{Fles92,Van93,Seno94}.

While there is little exact or analytical information available 
for this problem, a large amount 
of numerical work has been done using Monte-Carlo \cite{Dick84,Lam87,Lam88},
exact enumeration \cite{Chang88,Madr90,Fles92} and 
transfer matrix \cite{Derr83,Seno94} methods. 
The results of these analyses are controversial, however,
and it is this which triggered the work to be described here. 

In order to describe the purposes and the goals of this paper let us first
introduce some notation.

Let $n,b,s,k$ denote the number of sites, bonds, solvent interactions
and contact interactions of the lattice animal. We are working
on a square lattice with coordination number $\gamma=4$. Then the 
relation\footnote{In two dimensions, this is valid for $\gamma\leq 4$.}
$\gamma n=2b+2k+s$ implies that three of the four variables 
are needed to specify the problem completely. 
The statistical mechanics of the model is most conveniently
described by a generating function. This has recently been defined in
the literature \cite{Fles92,Seno94} in terms of two sets of variables 
and it is useful to give both of them here.

\BEA
{\cal Z} &=& \sum_{n} x^n {\cal Z}_n(y,\tau) = 
\sum_{n,b,k} a_{n,b,k} \, x^n y^b \tau^k \nonumber \\
         &=& \sum_{n} e^{\beta_0 n} {\cal Z}_n (\beta_1, \beta_2) 
= \sum_{n,s,k} a_{n,s,k} 
\exp \left( \beta_0 n + \beta_1 s + \beta_2 k \right)
\label{GenFun}
\EEA
where $a_{n,b,k}$ stands for the number of graphs with given $n,b,k$ and
$x,y,\tau$ and $\beta_0,\beta_1,\beta_2$ are fugacities.
The relationship between the two sets of variables
is given by 
\BEQ \label{generating}
x = \exp\left( \beta_0 + \gamma \beta_1 \right) \;\; , \;\;
y = \exp\left( - 2 \beta_1 \right) \;\; , \;\;
\tau = \exp\left( \beta_2 - 2 \beta_1 \right)
\EEQ
The infinite lattice animal can be described either by taking
$n\rar\infty$ or by fixing one of the three fugacities in terms of
the other two \cite{deGennes79}, 
say $x=x_c(y,\tau)$ or $\beta_0 = \beta_{0c}(\beta_1,\beta_2)$
(see also below). 

The compact and extended 
phases can be distinguished from the scaling
behaviour of the mean radius of gyration, $\langle R_N\rangle$, with
the number of monomers $N$, viz. 
\BEQ
\langle R_N\rangle \sim N^{\nu}.
\EEQ 
The phase diagram is sketched in Fig.~2.
First, there is a (second-order) transition from an extended
state (with $\nu \simeq 0.6408$ in two dimensions \cite{Derr82})
where the different branches of the lattice animal are widely
separated, to a compact state $(\nu =1/d)$. The locations of 
this extended--compact
transition which follow either from exact 
enumeration \cite{Fles92,Pear95} or 
transfer matrix calculations \cite{Derr82,Seno94} are in good 
agreement with each other. 
However, there is an important discrepancy, which is the subject of this
paper. While the enumeration studies \cite{Fles92,Pear95} 
suggest the existence
of an {\em additional} transition between two distinct compact phases,
no sign of such a transition was found 
using the transfer matrix \cite{Seno94}. 

The reason why a compact--compact transition could in principle be present 
can be better understood by looking at Fig.~1. 
Among the compact configurations (a,b,c)
which can fill all space, there are two different types. The first 
is a contact-rich (and almost linear) branched polymer (Fig.~1a,b) 
and the second 
type is a cycle-rich branched polymer (Fig.~1c). Contact-rich configurations 
are expected to provide 
the dominant contributions to $\cal Z$ in the $\tau\to\infty$ region,
whereas cycle-rich configurations are expected to dominate as $y\to\infty$.
The question is whether or not there exists a sharp transition between the 
compact configurations.\footnote{A similar kind
of transition has recently been discussed in the polymer 
literature \cite{Stel94,Cardy95} and it displays some analogies with
phenomena occurring in polymer gels, for which a multiplicity of phases can
be realized in the collapse process \cite{Annaka82}.}

Our aim in this paper is to further investigate 
on the existence or otherwise 
of this transition, by considering some approximations which 
render the model analytically solvable. This will be done using an exact
mapping onto an extended $q$--state Potts model, which will 
be defined in the next section. The results of the approximate analysis
of the extended $q$--state Potts model will then be reinterpreted in terms
of the lattice animal. Some of these results 
were already stated without derivation in \cite{Henk95}. 
In section~3, we present the exact solution of the extended Potts model
on the Bethe lattice and derive the phase diagram. This calculation will
be supplemented in section~4 by an analysis of the fixed-point structure
of the model resulting from 
the Migdal-Kadanoff renormalization group equations. 
In both schemes our results support a phase diagram with a single extended
and a single compact phase. These are separated by a line of 
$\theta$ transitions, divided into two sections which fall into different
universality classes and which are separated by a higher order 
multicritical point, which coincides with critical percolation. 
In section~5, we illustrate further the role of the competition of the
contact-rich and cycle-rich states by studying two-dimensional 
{\em directed} branched
polymers, where the cycles are absent. We find a single line of
$\theta$ transitions in the directed percolation universality class. 
Finally, in section~6 we present our conclusions. 

\section{Extended Potts model formulation}

We are interested in the phase diagram of infinite branched polymers
which are defined by the generating function $\cal Z$ in eqn~(\ref{GenFun}). 
Two existing results can serve as a guide. Firstly, in the case of
{\em strong embedding} (no contacts allowed), the extended-compact
transition is at $y\simeq 6.48$ and $\tau=0$ \cite{Derr83}. 
Secondly, it was shown
\cite{Fles92,Seno94} that along a certain line in the $x,y,\tau$ 
(or $\beta_0,\beta_1,\beta_2$) space the branched polymer can be mapped
onto percolation. The percolation critical point corresponds to the
following point on the $\theta$ transition line
\BEQ
y_p =2 \;\; , \;\; \tau_p =2 
\EEQ
(or $\beta_1=-\frac{1}{2}\ln 2, \beta_2=0$). 

Our approach is based on the known \cite{Giri77,HL81,Coni83,Wu82} 
exact relation of the lattice animal with the extended $q-$state Potts model
with the classical Hamiltonian
\BEQ \label{Potts}
{\cal H} = - J \sum_{(i,j)} \delta_{\sig_i,\sig_j} 
           - L \sum_{(i,j)} \delta_{\sig_i,1} \delta_{\sig_j,1}
           - H \sum_{i} \delta_{\sig_i,1}
\EEQ
where $\sig_i = 1,2,\ldots,q$ and the fugacities are expressed in terms
of $J,L,H$ as 
\BEQ
x = \exp[ -H - \gamma (J+L) ] \;\; , \;\;
y = \left( e^J -1 \right) e^{J+L} \;\; , \;\;
\tau = e^{J+L}
\EEQ
The relation of this extended 
Potts model with the lattice animal problem is 
\BEQ \label{Corres}
{\cal Z} = f_{\rm Potts} = 
\lim_{q\rar 1} \frac{\partial}{\partial q} \ln \widetilde{Z} \;\; , \;\;
\widetilde{Z} = \sum_{ \{ \sig \} } e^{-{\cal H} }
\EEQ
where $f_{\rm Potts}$ is the Potts model free energy. The
connection in eq.~(\ref{Corres}) 
can be established by considering the graphs in the
high-temperature expansion of the Hamiltonian (\ref{Potts}) and matching 
the weights in such a way 
as to reproduce the branched polymer generating function. 
Thus, calculating the phase diagram of the Potts model gives information
about the phases of the animal problem. {\em This correspondence is 
exact}. However, approximations are needed to treat the extended Potts 
model. Having solved the extended Potts model in the approximations
stated below, we shall reinterpret the results for the extended Potts model
phase diagram in terms of the lattice animal collapse transition(s). 
We shall pursue two approaches. 
\begin{enumerate}
\item The exact solution on the Bethe lattice should give the correct
topology of the phase diagram for sufficiently large spatial dimensions
$d$. While the Bethe lattice solution is expected to become exact in the
$d\rar\infty$ limit and should have the same universality properties
as a mean field treatment (see \cite{Baxt82} for a detailed discussion), 
it has the advantage over mean field theories that the self-consistency 
equations to be dealt with are considerably simpler. 

We point out that for the original branched polymer problem, a Bethe 
lattice solution is of little interest, since contact interactions are 
eliminated on the Bethe lattice by construction. Contact interactions may
be kept by considering the lattice animal on a cactus lattice, which is 
the dual lattice of the Bethe lattice, see \cite{Gujr95}. 
On the other hand,
contact interactions are not projected out 
by our treatment of the Bethe lattice approximation
in the extended Potts model, as we shall see below. 
\item We supplement this by an analysis of the fixed point structure of
the model, using the Migdal-Kadanoff approximation, which should become
reliable in $d=1+\eps$ dimensions, as reviewed in \cite{Bur82}. 
Our finding that the structure of the 
phase diagram is the same in {\em both}
approximations makes it plausible that the topology is valid for all
dimensions. This result is the more remarkable since it is known,
for example in three--component lattice gases \cite{Kauf81},
that the phase diagrams found from mean field
theory and Migdal-Kadanoff approximations may in general be distinct.
\end{enumerate}

Our aim is to find the phase diagram of {\em infinite} lattice animals
through a study of the equivalent extended Potts model. It is well known
that taking the limit
$n\rar\infty$ corresponds to finding critical points of the Hamiltonian
(\ref{Potts}). Transitions between different infinite polymer states
then correspond to multicritical points within the critical manifold
of the extended Potts model, see \cite{deGennes79}. 

We now recall a relationship \cite{Seno94} which 
will later be used to simplify the calculations on the Bethe lattice. 
In principle, we are interested in the lattice animal
free energy $F$. 
In the canonical ensemble (with $n$ fixed), it is rigorously 
known \cite{Fles92,Fles94} that
\BEQ \label{FleLim}
F(\beta_1,\beta_2) = \lim_{n\rar\infty} n^{-1} 
\ln {\cal Z}_n (\beta_1, \beta_2)
\EEQ
exists and is a convex function in $\beta_1$ and $\beta_2$. We now apply 
eq.~(\ref{FleLim}) to fix $x$ in the 
grand canonical ensemble such that we describe
the infinite lattice animal. Using eqn~(\ref{FleLim}), the animal generating
function takes the asymptotic form 
${\cal Z} \sim \sum_n ( x e^F )^n$, which diverges
at the critical point $x_c$ given by
\BEQ \label{Fxc}
F = - \ln x_c .
\EEQ
Thus, working in the grand canonical ensemble, it is sufficient to find
$x_c$ as a function $x_c(y,\tau$) of the other fugacities $y$ and $\tau$. 
Once the critical value of $x$ is obtained as a function 
$x_c(y,\tau)$, the canonical free energy follows from eqn~(\ref{Fxc}). 

\section{Bethe lattice solution}

We now describe in detail the exact solution of the $q\to 1$ extended 
Potts model on the Bethe lattice. 

\subsection{Generalities}

Consider a Cayley tree (Figure~3)
with $\gamma=3$ nearest neighbors for each site. 
Following the terminology of \cite{Baxt82}, by Bethe lattice solution we
mean the behaviour deep inside the Cayley tree. 
Let $\sig_0$ denote the central spin of the Bethe lattice.
Then the extended Potts model partition function is
\BEQ
\widetilde{Z} = \sum_{\sig_0} e^{H \delta_{\sig_0,1}} 
\sum_{\{s\}} \prod_{j=1}^{3} 
Q_n \left( \sig_0 \left| s^{(j)} \right. \right) 
\EEQ
where $s^{(j)}$ denotes the spins in the j$^{th}$ sub-branch and
\BEA
Q_n \left( \sig_0 \left| s \right. \right) &=& 
\exp \left( J \delta_{\sig_0, s_1 } + L \delta_{\sig_0,1}\delta_{s_1,1}
+H \delta_{\sig_0,1} \right) \nonumber \\
&\cdot& \exp\left( J {\sum_{(i,j)}}' \delta_{s_i,s_j}
+ L {\sum_{(i,j)}}' \delta_{s_i,1}\delta_{s_j,1}
+ H {\sum_i}' \delta_{s_i,1} \right)
\EEA
where ${\sum}'$ is a sum over one sub-branch 
of which the first site is $s_1$.

\subsection{Derivation of the self-consistency equations}

We define, following Baxter \cite{Baxt82} 
\BEQ
g_n(\sig_0 ) = { \sum_{ \{ s \} } }' Q_n \left( \sig_0 | s \right)
\EEQ
where $n$ refers to the number of iterations performed in the construction
of the Cayley tree. Recursion relations for the $g_n$ are
\BEQ \label{GRec}
g_n(\sig_0) = \sum_{s_1=1}^{q} 
\exp\left( J \delta_{\sig_0,s_1} + L \delta_{\sig_0,1}
\delta_{s_1,1} + H \delta_{s_1,1} \right) 
\left( g_{n-1}(s_1) \right)^{2} . 
\EEQ
Having solved eqs.~(\ref{GRec}) and thus obtained the $g_n$, the
thermodynamics is determined completely. 

The analysis of the eqs.~(\ref{GRec}) can be simplified, in analogy to the
procedure described in \cite{Bara83}, by introducing the variables
\BEQ \label{Thermo}
\Xi_n = \frac{ g_n(\sig_0 \neq 1,2,3) }{g_n (1) } \;\; , \;\; 
\ups_n = \frac{ g_n (2) }{ g_n (1) } \;\; , \;\;
Z_n = \frac{ g_n (3) }{ g_n (1) }
\EEQ
which, as will become apparent later, are sufficient to describe
the full thermodynamics. Our choice of variables was motivated by the
mean field treatment of the ordinary $q$--states Potts model 
in an external symmetry--breaking field \cite{Bara83}. 
There it was shown that for the distinct ground states
corresponding to different phases the order parameter either has the
same value for all its $q$ components or that at most the order parameter
component which is directly coupled to the field may be different from 
the others. In (\ref{Thermo}) we generalize that result in order to allow 
for the possibility of a compact--compact transition besides the
expected $\theta$ transition.  
As $n\rar\infty$, the variables in (\ref{Thermo}) tend to fixed point values
$\Xi, \ups, Z$, which can be determined from the recursion relations 
(\ref{GRec}). At this stage, we take the $q\rar 1$ limit 
(see (\ref{Corres})) and find the self-consistency relations 
\BEA \label{SelfCon}
\Xi &=& \frac{ x^{-1}\tau^{-3} + \ups^2 + Z^2 + (y/\tau -2) \Xi^2}
{x^{-1}\tau^{-2} + \ups^2 + Z^2 - 2 \Xi^2 } ,\nonumber \\
\ups &=& \frac{ x^{-1}\tau^{-3} + ( y/\tau +1)\ups^2 + Z^2 - 2 \Xi^2}
{x^{-1}\tau^{-2} + \ups^2 + Z^2 - 2 \Xi^2 } ,\\ 
Z &=& \frac{ x^{-1}\tau^{-3} + \ups^2 + (y/\tau+1) Z^2 - 2 \Xi^2 }
{x^{-1}\tau^{-2} + \ups^2 + Z^2 - 2 \Xi^2 } .\nonumber
\EEA
The equations (\ref{SelfCon}) can be decoupled after some transformations
of the variables. We then find that there are four cases 
which must be considered separately. 
As we now show, in each case the problem of solving a set of three coupled
equations can be reduced to the solution of a single quartic equation. 

\begin{enumerate}

\item {\bf Case A: $\Xi=\ups=Z$}. \\ The fixed point equation is a quadratic
equation in $\Xi$, with the solutions 
\BEQ \label{ASol}
\Xi_{\pm} = \frac{ 1 \pm \sqrt{ 1 - 4 x y } }{ 2 x y \tau}
\EEQ
{}From eqn~(\ref{SelfCon}), it is clear that $\Xi_{-}$ is the stable solution 
which, using eqn~(\ref{Corres}), determines the lattice 
animal generating function
\BEQ
{\cal Z} = x \tau^3 \Xi_{-}^3
\EEQ
These solutions are real provided $4 x y \leq 1$. Criticality corresponding
to an infinite lattice animal is recovered at 
\BEQ \label{ACrit}
x_c = \frac{1}{4 y}.
\EEQ

\item {\bf Case B: $\Xi=Z\neq \ups$}. \\ We introduce the variables
\BEQ \label{uv}
u = \Xi + \ups \;\; , \;\; v = \ups - \Xi.
\EEQ
For $v=0$, case A is recovered. For $v \neq 0$, a factor $v$ cancels
from the fixed point equations (\ref{SelfCon}) giving 
\BEQ
v = \frac{y}{\tau} - \frac{1}{x \tau^2} \frac{1}{u},
\EEQ
\BEQ \label{QB}
x^2 y \tau^4 \left( u^4 - 4 u^3 \right) 
+ x \tau^2 \left( 4 (\tau-1) - x y^3 \right)
u^2 + 2 x y^2 \tau u - y  = 0.
\EEQ

\item {\bf Case C: $\Xi\neq \ups = Z$}. \\ We introduce again $u$ and $v$
defined by (\ref{uv}). For $v=0$, case A is again recovered. 
For $v\neq 0$, a factor $v$ cancels giving
\BEQ
v = \frac{y}{2\tau} - \frac{1}{2x \tau^2} \frac{1}{u},
\EEQ
\BEQ \label{QC} 
x^2 y \tau^4 \left( u^4 - 4 u^3 \right) + 
x \tau^2 \left( 4 (\tau-1) - \frac{1}{4}x y^3 \right)
u^2 + \frac{1}{2} x y^2 \tau u - \frac{y}{4}  = 0.
\EEQ

\item {\bf Case D: $\Xi\neq\ups\neq Z$}. \\ We define new variables
\BEQ
a = \ups + Z \;\; , \;\; b = \ups - Z
\EEQ
and rewrite the fixed point equations (\ref{SelfCon}) 
in terms of $\Xi,a,b$. $b=0$ corresponds to case C. 
For $b\neq 0$, a factor $b$ cancels giving
\BEQ
b^2 = - a^2 + 4 \Xi^2 + \frac{2 y}{\tau} a - \frac{2}{x \tau^2}
\EEQ
together with
\BEA
\frac{y}{\tau} a^2 &=& \frac{2}{x\tau^3} - 4\Xi^2 + (y/\tau+2) 
\left( \frac{y}{\tau} a + 2 \Xi^2 -\frac{1}{x\tau^2} \right),
\nonumber \\
\frac{y}{\tau} a \Xi &=& \frac{1}{x \tau^3} +\frac{y}{\tau} \Xi^2
+\frac{y}{\tau} a - \frac{1}{x\tau^2}. \label{axieqs}
\EEA
Provided $\Xi \neq 1$, we solve the second of eqs.~(\ref{axieqs}) for $a$
\BEQ
\frac{y}{\tau}\left( \Xi -1\right) a = \frac{y}{\tau} \Xi^2 
+\frac{1-\tau}{x\tau^3}
\EEQ
and find
\BEA \label{QD}
\lefteqn{ x^2 y^2 \tau^4 \left( \Xi^4 + (y/\tau-2) \Xi^3 \right)
- x y^2 \tau^2 (1+x y \tau^3 ) \Xi^2 } \nonumber \\
&+& x y \tau ( y(1+\tau) +2\tau(1-\tau))\Xi
- (1-\tau)^2 - x y^2 \tau =0
\EEA
Note that the relation between $\Xi$ and $Z$ remains unspecified. 
\end{enumerate}

\subsection{Analysis of the self-consistency equations}

The problem has been reduced to the solution of three quartic equations
(\ref{QB}),(\ref{QC}) and (\ref{QD}). 
Further analysis is greatly simplified
by numerically checking that all solutions in 
Case D where $\Xi$, $\ups$ and $Z$ are 
{\it a priori}  different, reduce to one of case B, with $\Xi=Z$. 
As we shall show below, this implies that there
are only two distinct phases for the infinite lattice animal on the
Bethe lattice. In particular, that means if any two branches of the lattice
animal generating function calculated in the cases A,B,C meet, all three
solutions have to coincide. It follows that the topology of the phase
diagram calculated in the whole $(\Xi,\ups,Z)$--space is the same as would
have been found when only considering $(\Xi,\ups)$--space, with $\ups=Z$.
We expect the same reduction to occur in even more generic setups than
done in eq.~(\ref{Thermo}).

We only have to consider the cases A,B,C.  
We introduce a new variable
\BEQ
p= u \tau \;\; , \;\; u = \Xi + \ups. 
\EEQ
We recall from (\ref{Fxc}) that in order to get the infinite lattice 
animal free energy in the canonical ensemble, it is enough to know $x$ as 
a function of $y$ and $\tau$. Solving for $x=x_c(p;y,\tau)$ rather than 
solving for $p$ (or $u$) has the further advantage of going from quartic 
to quadratic equations. $p$ now plays the role of an order parameter and 
will be fixed by maximizing $x(p)$ with respect to $p$. We find
\begin{enumerate}

\item {\bf Case A}. We have
\BEQ
A(p;y,\tau) := x^{(A)} (p;y,\tau) = \frac{2(p-2)}{y p^2}
\EEQ

\item {\bf Case B}. The equation is
\BEQ
x^2 ( y p^4 - 4 y \tau p^3) 
+ x \left( 4(\tau-1)p^2 - x y^3 p^2 + 2 y^2 p\right) - y =0
\EEQ
with the solutions
\BEA
B_{\pm}(p;y,\tau) &:=& x_{\pm}^{(B)}(p;y,\tau) \\
&=& \frac{2(1-\tau)p-y^2}{py(p^2-4\tau p-y^2)}
\pm \frac{\sqrt{p(4\tau^2-8\tau+4+y^2)-4y^2}}
{\sqrt{p} (p^2-4 \tau p-y^2) y}. \nonumber
\EEA

\item {\bf Case C}. The equation is
\BEQ
x^2 ( y p^4 - 4 y \tau p^3) + 
x \left( 4(\tau-1)p^2 - \frac{1}{4}x y^3 p^2 + \frac{1}{2} y^2 q\right) 
- \frac{y}{4} =0
\EEQ
with the solutions
\BEQ
C_{\pm}(p;y,\tau) := x_{\pm}^{(C)}(p;y,\tau) = 
\frac{1}{2} B_{\pm}(p;y/2,\tau).
\EEQ
\end{enumerate}

To get the equilibrium physics, we have to minimize the lattice animal
free energy $F(y,\tau)$ (see eqs.~(\ref{FleLim},\ref{Fxc})) with respect 
to $p$ or equivalently to maximize $x(p;y,\tau)$. For case A, we have
\BEQ
F = -\ln x = - \ln \left( \frac{2(p-2)}{p^2 y} \right) \;\; , \;\;
\frac{\partial F}{\partial p}= \frac{p-4}{p(p-2)} \;\; , \;\;
\left. \frac{\partial^2 F}{\partial p^2}\right|_{p=4} = \frac{1}{8} > 0
\EEQ
and $F$ has a single minimum at $p=4$. Indeed, $F=\ln (4 y)$ evaluated
at $p=4$ is concave in $y$. 

It remains to be seen under which conditions this local minimum of 
$F^{(A)}$ is the absolute minimum\footnote{It is enough to consider
$B_{-}$ and $C_{-}$, since the other two solutions $B_{+}, C_{+}$ have
no maxima with respect to $p$.} of $F$. 
>From the above it is clear that if any two of the solutions A,B,C
meet, we have in fact a simultaneous meeting of all three of them. 
This common meeting point occurs at $p=4$
\BEA
A(4;y,\tau) = B_{-}(4;y,\tau)=C_{-}(4;y,\tau) = \frac{1}{4 y} \;\; ; \;\;
\tau \geq 1 , \nonumber \\
A(4;y,\tau) = B_{+}(4;y,\tau)=C_{+}(4;y,\tau) = \frac{1}{4 y} \;\; ; \;\;
\tau \leq 1 .
\EEA
In addition, we have checked that
\BEQ
C_{-}(p;,y,\tau) \geq B_{-}(p;y,\tau) .
\EEQ 
For an example illustrating the solutions $A,B_-,C_-$, see Figure~5. 
This implies that there are just two distinct phases of the 
lattice animal which are described
by the two solutions $A(p;y,\tau)$ and $C_{-}(p;y,\tau)$. Furthermore,
we want solutions such that the fugacity $x$ is real and positive. This
yields the condition
\BEQ
p \geq \left\{ \begin{array}{lr} 2 & \mbox{\rm Case A} \\
\frac{(y/2)^2}{(\tau-1)^2+(y/4)^2} & \mbox{Case C} \end{array}
\right. 
\EEQ
On the other hand, the unstable solution $\Xi_+$ 
from case A always corresponds
to $p\geq 4$. Furthermore, we argue in the appendix that 
for $p>4$ the solution
$C_-$ is not a stable solution of the self-consistency 
relations (\ref{SelfCon}). We therefore must have
\BEQ
p \leq 4
\EEQ

To get the
transition lines between the two phases, note that
$\partial C_{-}/\partial p =0$ if and only if $\tau=2$.
Furthermore
\BEQ
C_{-}(p;y,2) ~ ~ \mbox{\rm  has at $p=4$ a } \left\{ \begin{array}{l}
\mbox{\rm maximum if $y>8$} \\
\mbox{\rm turning point if $y=8$} \\
\mbox{\rm minimum if $y< 8$} \end{array} \right. 
\EEQ 
To locate the transitions, one must distinguish two cases. First, for
$y<8$, one has a first order transition. It is given by the conditions
\BEA
C_{-}(p;y,\tau) &=& A(4;y,\tau) = \frac{1}{4 y} \nonumber \\
\frac{\partial C_{-}}{\partial p} (p;y,\tau) &=& 0 \\
\frac{\partial^2 C_{-}}{\partial p^2} &\leq& 0 \nonumber 
\EEA
At this transition point, $p$ jumps from its value $p=4$ for $\tau$ small
to a new value $p_c(y,\tau)<4$. The location of the line and $p_c$ have
to be found numerically. 

Second, for $y>8$ and $\tau>2$, 
numerical studies show that $C_{-}(p;y,\tau)$
has a maximum at some value $p' <4$. For all $p<4$, $C_{-} > A$ and the
state described by $p'$ is the equilibrium ground state. As $p\rar 4$, the
two solutions approach each other continuously. 

Finally, for $y=8$ and $\tau=2$, the second order line ends in
a tricritical point. The two transitions
are shown in Fig.~6. We also show the jump 
in $p$, $\Delta p = 4-p_c(y,\tau)$
and observe that $\Delta p \sim (8-y)$ for $y\to 8^-$. 

\subsection{Summary}

Studying the phase diagram of the extended 
Potts model in the $q\rar 1$ limit
on the Bethe lattice, we find that there are two distinct phases
(A and C above) on the manifold corresponding to the infinite
lattice animal. The transition between them is first order for
$y<8$ and second order for $y>8$. There is a tricritical point at 
$y=8, \tau=2$. The Bethe lattice calculation corresponds to a mean-field 
calculation \cite{Baxt82}. Detailed comparison \cite{Baxt82} of the exact 
solution of the simple $q-$ states Potts model (that is, $L=H=0$) in two 
dimensions with the Bethe lattice solution suggests that the latter
faithfully represents the model behaviour for infinite dimensionality. 
We expect the same for the more general models treated here.

\section{Migdal-Kadanoff renormalization group}

In the last section, we studied the extended Potts model (\ref{Potts})
using an approximation which should become more reliable with increasing
dimensionalty. We now wish to complement this calculation by a
Migdal-Kadanoff renormalization group study. The
Migdal-Kadanoff approximation preserves the self-duality property of
several two-dimensional models (in particular the 
$q-$state Potts model, see \cite{Bur82,Itz89}) and thus yields the
phase diagram exactly. The bond-moving approximation involved is exact
at zero temperature and its
predictions for exponents are exact to order $\eps$ in $d=1+\eps$ 
dimensions, as reviewed in detail in \cite{Bur82}. 
We therefore hope that the renormalization group calculation
will provide reliable information on the fixed point structure for
low dimensions.

The Hamiltonian (\ref{Potts}) has already been investigated using
a Migdal-Kadanoff renormalization group by Coniglio \cite{Coni83}. 
He used a rescaling factor $b=2$ in dimensionality $d=2$ and considered
$q=1$, rather than $q\to 1^+$. He identified four non-trivial fixed
points which correspond to (1) an extended lattice animal phase,
(2) a compact phase, (3) a percolation point and (4) a tricritical point
analogous to a $\theta$ point. These are not sufficient to describe the
topology of Fig.~2 and therefore it seemed useful to look again at the
calculation with particular emphasis on the possibility of the occurrence
of a compact--compact transition. 

Here, we shall consider the case $d=1+\eps$. Furthermore, as will become
apparent, the limit $q\to 1$ has to be taken with care, 
because the fixed point structure changes at $q=1$. 
For $d=1+\eps$ and $b=2$ the Migdal-Kadanoff recursion equations for the
extended Potts Hamiltonian (\ref{Potts}) are 
\BEA \label{Recursion1}
\xi^{'} &=&
\left( 
{{\xi(1+\rho+(q-2)\eta)}\over{1+\xi^2+(q-2)\eta^2}}
\right)^{b^{\eps}}
\\
\label{Recursion2}
\eta^{'}&=&
\left(
{{\xi^2+2\eta+(q-3)\eta^2}\over{1+\xi^2+(q-2)\eta^2}}
\right)^{b^\eps}
\\
\label{Recursion3}
\rho^{'}&=&
\left( 
{{\rho^2+(q-1)\xi^2}\over{1+\xi^2+(q-2)\eta^2}}
\right)^{b^\eps} 
\EEA
where
\BEQ
\xi = \exp(H/2 -J) \;\; , \;\; \eta = \exp(-J) \;\; , \;\;
\rho = \exp(L+H)
\EEQ
Eqs. (\ref{Recursion1})-(\ref{Recursion3}) were obtained by performing a 
one-dimensional decimation followed by bond-moving
\cite{Bur82,Kauf81}. 

The fixed point structure that follows from the recursion
equations (\ref{Recursion1}-\ref{Recursion3}) is complicated and 
$q$-dependent. However, using the $q=2,3$ cases as a guide \cite{Kauf81},
in the two limits of interest to us ($q \to 1$ and
$\eps \to 0$) a clear pattern appears and 14
fixed points can be identified. These are listed in Table~\ref{tab2}, 
together with the
eigenvalues of the Jacobian evaluated at the fixed points. We also
illustrate the fixed point structure in Fig.~6.

Note that that the fixed
points naturally divide into three groups. The first group contains the
fixed points B, C and D for which $\rho^* =\infty$. The second group, which
is the one we shall be interested in, contains the fixed points 
E,F,G,H and I which all have $\rho^* \simeq 1$. Finally, the third group
contains the fixed points with $\rho^* \approx 0$, namely A
and the cluster of fixed points around $y\simeq 1$, J,K,L,M and N. For
$q\to 1$, the last five points merge to a single fixed point.

To understand the physics, we begin by singling 
out the completely attractive (or trivial) fixed points. Reading from 
table~\ref{tab2} the fixed points B,C,D and A,J,N only have eigenvalues 
which are smaller than one. Hence they cannot describe
the critical behaviour of the infinite branched polymer.
Next, we consider the cluster of five fixed points near to 
$(\xi^*,\eta^*,\rho^*)\approx (0,1,0)$. One (M) has two relevant
eigenvalues, whilst two (K and L) have one relevant eigenvalue,
with an associated exponent $y=d$. From the eigenvectors, we have
checked that the renormalization group flow from the vicinity of
one of the fixed points M,K,L always goes towards the trivial stable
fixed points J and N. Since these five fixed are 
going to merge into a single
fixed point in the $q\to 1$ limit and none of them has three
relevant eigenvalues as would be required from a percolation fixed point, 
we conclude that they cannot govern
any part of the phase diagram of the branched polymer problem. 
We thus find that the fixed point structure depends on carefully taking
the $q\rar 1^{+}$ limit, rather than simply setting $q=1$. 

It remains to consider those fixed points which are close to the
$\rho^*=1$ plane. Two of these (E and I) are independent of both
$q$ and $\eps$ and have one relevant eigenvalue. 
For the fixed point E, the relevant direction is characterized
by the exponent $1/\nu =d =1+\eps$. This is characteristic
of a compact object and we therefore conclude that this fixed point controls
the compact phase of the infinite branched polymer. 
For the fixed point I, the single relevant eigenvalue is
$1/\nu = d-1$. Since $\nu > 1/d$, we expect this to describe 
a non-compact phase and the fixed point I to govern the extended
phase of the polymer. 

The fixed points F, G and H are dependent on $q$ and $\eps$, but
for $q < 2$ all appear in the plane $\rho=1$, to 
leading order in $\eps$. As 
$d$ decreases they move towards the point E. 
The fixed point H has three relevant eigenvalues and therefore
represents the percolation fixed point (which is realized for
$H=0$ and $L=0$ in (\ref{Potts}) and for $d=2$ the Migdal-Kadanoff approach
is known \cite{Bur82,Itz89} to correctly reproduce the exact critical
point). F and G are tricritical points
and govern the renormalization group flow along the two critical lines 
leaving the percolation point H. We point out that the fixed point G
is only found when the limit $q\to 1^+$ is carefully taken. However,
if one simply puts $q=1$, that fixed point is missed, in agreement with
the earlier work of Coniglio \cite{Coni83}.

Thus we predict a line of $\theta$ points which contains two segments
falling into two
distinct universality classes and meeting at the percolation point. 
The $\theta$ line separates the extended phase from a single compact phase.

\section{Directed lattice animals}

So far, in this paper we have considered the structure of the phase diagram
of the infinite branched polymer. Although the interplay of the
different ground states (see Figure~1) 
in the contact-rich and contact-poor region
of the compact phase might at first sight suggest the existence of a
compact-compact transition, this was not
supported by our (albeit approximate) calculations. To provide some
insight into the role of the different ground states, we now wish to
consider a model where the spiralling or folding ground 
states shown in Fig.~1a,b are eliminated. 

For this reason, we now consider
{\em directed} branched polymers. This model is defined in complete analogy
with the system discussed so far, but requiring that the bonds, solvents
and contacts are directed. We use a square lattice with the preferred
direction along a lattice diagonal. Animal configurations 
are allowed only if 
(1) they start from a single site (2) all bonds have a component in a
preferred direction and (3) the solvents and contacts counted in
the generating function have a component in the preferred direction. 
The special case where contacts are forbidden (that is, the limit
$\beta_2 \rar -\infty$) was studied previously by Dhar \cite{Dhar87}. 

The interest in this model arises because due to the directedness
condition ground states of the {\em undirected} lattice animals in the 
contact-rich region ($\beta_1\rar -\infty, \beta_2\rar\infty$) are not 
allowed. Indeed, the phase diagram 
of the directed model (Fig.~7) is quite different, as we shall now show.

For directed lattice animals, it is always possible to reduce the 
problem to the ``strong-embedding'' ($\beta_2\rar -\infty$)
case. To see this, consider a strongly-embedded lattice animal
(where all neighboring sites are connected by bonds). Strong embedding
is broken when some of the bonds are replaced by contacts. We now ask
how often these replacements of bonds by contacts can be made. 

Because of the directedness, it is clear that bonds 
can be replaced by contacts
only if the strongly-embedded lattice animal contains closed loops. 
Furthermore, each loop contains exactly one ``final'' site. Only one
of the two bonds leading to this final
site may be replaced by a contact because of condition (1). Now, consider
a directed lattice animal with $c$ loops and denote by $g(c,k)$ the 
number of ways of replacing exactly $k$ of the bonds by contacts. 
Obviously, $g(c,k)=0$ if $k>c$, since no loop can be cut twice. It is
also clear by inspection that $g(c,0)=1$ and $g(c,1)=2c$. Also, 
for $k\geq 2$ the recursion formula
\BEQ \label{gRek}
g(c,k) = 2 g(c-1,k-1) + g(c-1,k)
\EEQ
is valid. To see this, imagine placing the first contact onto the directed
lattice animal. Single out one of the loops. Either this contact cuts this
loop, which can be done in exactly two ways, and one then has to place
$k-1$ contacts onto an animal with $c-1$ loops or this loop is not cut at all
and all $k$ contacts are placed onto the remaining $c-1$ loops. 
Defining the function
\BEQ
f(c;\beta) = \sum_{k=0}^{c} g(c,k) e^{\beta k}
\EEQ
it is easy using (\ref{gRek}) to show that for all $c\geq 0$
\BEQ \label{fRek}
f(c;\beta) = \left( 1 + 2 e^{\beta} \right) f(c-1;\beta) =
\left( 1+2 e^{\beta} \right)^c
\EEQ
Consider now the directed lattice animal generating function. On the 
square lattice, $2n = b+k+s$. Using the Euler relation
$c=b-n+1$ and (\ref{fRek}) 
\BEA
{\cal Z} &=& \sum_{\cal G} a(n,k,s) e^{\beta_0 n} e^{\beta_1 s} e^{\beta_2 k}
\nonumber \\
&=& {\sum_{\overline{{\cal G}}} }^{'} a(n,0,s) e^{\beta_0 n} e^{\beta_1 s}
\left( 1 + 2 e^{\beta_2} \right)^c 
\EEA
where $a(n,k,s)$ is the number
of directed lattice animals with $n$ sites, $k$ contacts and $s$ solvents.
$\cal G$ runs over all directed lattice animals and
$\overline{{\cal G}}$ over all strongly embedded directed lattice
animals. In the strong-embedding case, $k=0$. Thus $c=n-s+1$ and 
\BEQ
{\cal Z} = \left( 1 + 2 e^{\beta_2} \right)
{\sum_{\overline{{\cal G}}} }^{'} a(n,0,s) 
\left( e^{\beta_0} \left( 1+ 2 e^{\beta_2}\right) \right)^n
\left( \frac{ e^{\beta_1} }{1+2 e^{\beta_2} } \right)^s 
\EEQ
This is the strongly embedded
directed lattice animal, with the effective solvent fugacity
\BEQ
e^{\widetilde{\beta_1}} =  \frac{ e^{\beta_1} }{1+2 e^{\beta_2} }
\EEQ
The strong-embedding model can be mapped onto directed percolation 
\cite{Dhar87}. We have thus shown that the fully directed lattice animal
has a single extended-compact transition which is in the directed percolation
universality class\footnote{We have also checked that the relationship
with directed percolation can be established directly from the fugacities
given to the elementary processes of the directed animal model.}. 
In contrast to the undirected model, there is
no change of the values of the critical exponents along the transition line. 

\section{Conclusions}

In this paper, we have investigated the phase diagram of the extended
Potts model (\ref{Potts}) in the $q\to 1$ limit. We have concentrated
on the structure of the manifold of critical points, because this is
related to the phase diagram of an infinite branched polymer. Using
Bethe lattice and Migdal-Kadanoff renormalization group calculations, 
we find a line of $\theta$ transitions between the extended and the 
compact phase. The $\theta$ line consists of two segments which are 
controlled by different fixed points. Their meeting point coincides
with the percolation critical point. Although these results were
obtained from rather drastic approximations, we expect the results
coming from these to
be reliable for large (Bethe lattice, $d\to\infty$) or small
(Migdal-Kadanoff, $d=1+\eps$) dimensions, respectively. Furthermore,
since the qualitative structure of the phase diagram is the same in both
cases, it is plausible that the topology of the phase diagram is 
indeed independent of the dimensionality. Our results on the structure
of the $\theta$ line agree with previous
numerical calculations using either graph enumeration or transfer matrix
techniques. 

Our calculations give no evidence
for a further transition between two compact phases. While that conclusion
does agree with numerical transfer matrix calculations, it is in 
disagreement with analysis based on graph enumeration. We thus expect
the crossing--over between the contact-rich ground states Figure~1ab and
the bond-rich ground state Figure~1c to be smooth. While intuitively a phase
transition between these ground states might appear plausible at least in two 
dimensions because a ground state as the one in Figure~1a could be seen
as a vortex while the state in Figure~1c is not, 
the nature of that transition is less apparent in three and
higher dimensions. However, graph enumeration predicts \cite{Pear95} a
compact--compact transition in three and more dimensions as well.

This does not mean, however, that the different ground states of the model
in different regions of the compact phase have no physical role. This
is clearly illustrated by the case of {\em directed} branched polymers,
where the spiralling states (Figure~1ab) are absent. Then the multicritical
point on the $\theta$ line found for the {\em undirected} branched 
polymer disappears and the entire $\theta$ line is in the directed
percolation universality class.

\zeile{3} \noindent
{\bf Acknowledgements:} It is a pleasure to thank S. Flesia, A.L. Stella,
P. Peard, S.G. Whittington and J.M. Yeomans for useful discussions. 
We thank P. Peard for kindly providing his numerical results before 
publication. MH thanks the Dipartimento di Fisica della
Universit\`a di Padova, where part of this work was done, 
for hospitality and FS thanks the Sub-Department of Theoretical 
Physics in Oxford where another part was added. 
MH was supported by a EC grant of the ``Human Capital and Mobility''
program. 
 
\appsection
Here we study the stability of the solutions of the self-consistency 
equations (\ref{SelfCon}). These arise from recursions of the form
\BEQ
x_{n+1} = f( x_n) 
\EEQ
and it is well known that a fixed point solution 
$x^* = \lim_{n\rar\infty} x_n$
is stable under small perturbations if and only if $|f'(x^*)|<1$. 
Consider first case A. Then we have the recursion 
\BEQ
\Xi_{n+1} = f(\Xi_n) = \frac{1}{\tau} + x y \tau \, \Xi_n^2
\EEQ
With the solutions $\Xi_{\pm}$ from eq.~(\ref{ASol}) we have
\BEQ
f'(\Xi_{\pm}) = 1 \pm \sqrt{ 1 - 4 x y} 
\EEQ
and thus $\Xi_{-}$ is the stable and $\Xi_{+}$ the unstable fixed point. 
Since $p_{\pm} = 2 \tau \Xi_{\pm}$, it follows that if $\tau\simeq 2$ as
is the case for the A/C transition, we have 
$p \geq 4$ for the unstable case. Next, we look at
case C. We introduce $u,v$ from (\ref{uv}) and are interested in the case
when $v$ is small, that is, near to the $\theta$ transition. 
Then the following
fixed point equation for $u$ holds
\BEQ
u \simeq \frac{2}{\tau} + \frac{1}{2} x y \tau u^2 + {\cal O}(uv) 
\EEQ
which the same relation as for $2 \Xi$. From (\ref{ASol}) it then follows
that if $u \geq u_{+} > 1/(x y \tau)$, it is an unstable solution. On
the other hand, in the vicinity of the A/C transition we have from 
(\ref{ACrit}) that at least in the transition region, all solutions 
which correspond to $p = \tau u > 1/(xy) \simeq 4$, will be unstable. They 
are eliminated by requiring that $p \leq 4$.  

\newpage

\begin{table}
\begin{tabular}{|c|ccc|ccc|c|} \hline
 & $\xi$ & $\eta$ & $\rho$ & 
\multicolumn{3}{c|}{Eigenvalues}& Comments \\ \hline
A & 0 & 0 & 0 &   0 & 0 & 0  &\\
B & 0 & 1 & $\infty$ &   0 & 0 & 0  &\\
C & $\infty$ & 1 & $\infty$ &   0 & 0 & 0  &\\
D & 0 & 0 & $\infty$ &   0 & 0 & 0  &\\
E & 0 & 0 & 1 &  0 & 0 & $2^d$  &\\
F & $\beta$ & 0 & 1 &  0.738 & 1.23 & 2.014 &  $\eps=0.01$ \\ 
  &         &   &   &  0.736 & 1.213 & 2  & $\eps\to 0$ \\
G & 0 & $\beta$ & 1 &  0 & 1.22  & 2.014 & $\eps=0.01$ \\
  &   &         &   &  0 & 1.213 & 2  & $\eps\to 0$ \\
H & $\alpha$ & $\alpha$ & 1  & 1.0045 & 1.0045 & 2.014 & $\eps=0.01$ \\
  &          &          &    & $1+$ & $1+$ & 2  & $\eps\to 0$ \\
I & 1 & 1 & 1 &  0 & 0 & $2^{d-1}$  &\\
J & 0 & 1 & 0 &  0 & 0 & 0  &\\
K & 0 & 1 & $(q-1)^{e/(2e-1)}$ &  0 & 0 & $2^d$  & \\
L & 0 &$\gamma$ & $\delta$ & 0 & $\lambda_1$ & $2^d$ &  \\
M & $1.19 \cdot 10^{-8}$ & 1 & $0.010$ &  0 & $1.97$ & $2.002$ &
$q=1.01$, $e =1.001$ \\
N & 0 & $\gamma$ & 0 &  0 & 0 & $\lambda_2$ & \\ \hline
\end{tabular}
\caption[MKRG fixed points]{Fixed points of the Migdal-Kadanoff recursion
relations. The following abbreviations are used:
$\alpha=\frac{1}{2}\exp(-1/\eps)$, 
$\beta=\exp\left(-\frac{1}{\eps2\ln 2}\right)$,
$\gamma=\frac{1}{2-q}+\frac{\ln(2-q)}{q-2}\left(e-1\right)$,
$\delta=\frac{q-1}{q-2}+\frac{e-1}{q-2}
\left[ (q-1)\ln\left(\frac{q-2}{q-1}\right)-2\ln(2-q) \right]$,
$e=2^\eps$,
$\lambda_1=\frac{(q-2)\ln(2-q)}{q-1}\left(2^d -2\right)$,
$\lambda_2=\frac{2(q-2)}{q-1}(e-1)\ln(2-q)$. Analytic expressions
are correct to leading
order in $\eps$ or to the given order in $e-1$. 
The numerical eigenvalues for the fixed points F,G,H
are correct up to terms ${\cal O}(q-1)^2$. 
\label{tab2} }
\end{table}

\newpage

\noindent {\Large \bf Figure captions}
\zeile{1}
\noindent {\bf Figure~1}: Possible ground states of the infinite branched
polymer in (a) (b) the contact-rich region of the compact phase,
(c) the bond-rich region of the compact phase (d)
the extended phase. \\
\noindent {\bf Figure~2}: Numerical phase diagram of the infinite branched 
polymer in two dimensions in the variables (a) $\beta_1 - \beta_2$ 
(b) $y - \tau$ which are defined in the text. The extended phase is indicated
by E and the compact phase by C. 
Results are from cluster enumeration (circles)
\cite{Fles92,Pear95} and the transfer matrix (open squares) \cite{Seno94}. 
The full circles give the extended--compact collapse transition while the
open circles give the compact--compact transition predicted in 
\cite{Fles92,Pear95}. The black squares give the exactly known transition
at the percolation point \cite{Fles92,Seno94} and the collapse transition
in the strong embedding limit \cite{Derr83}. 
\\
\noindent {\bf Figure~3}: Bethe lattice with coordination number $\gamma=3$
and central spin $\sigma_0$. \\
\noindent {\bf Figure~4}: $p$-dependence of the three solutions
$A$ (dotted), $B_-$ (dashed) and $C_-$ (full) 
of the Bethe lattice recursion relations, for $y=2, \tau=1.61949$.  \\
\noindent {\bf Figure~5}: Phase diagram of the extended Potts model on 
the Bethe lattice. The phase labelled A corresponds to the extend lattice 
animal while C corresponds to the compact lattice animal. The line indicated 
by short dashes is first order and the full line is second order. The black 
square at $y=8,\tau=2$ indicates their meeting point. 
The jump $\Delta p$ as a function of $y$ 
between the phases A and C along the first order line
is indicated by the long-dashed line and is shown on the same scale. \\
\noindent {\bf Figure~6}: Fixed point structure which follows from
the Migdal-Kadanoff renormalization group 
for the extended Potts model (\ref{Potts}) for $q\to 1$ and $\eps\to 0$. 
The plane $\rho=1$ is also shown to guide the eye. \\
\noindent {\bf Figure~7}: Phase diagram for the directed branched polymer. 
The extended phase is indicated by E and the compact phase by C. \\


\newpage 
  
\begin{references}

\centerline{\bf References}

\bibitem{Flory} P. Flory, {\it Principles of Polymer Chemistry}, Cornell
University Press, (Ithaca NY, 1971)

\bibitem{deGennes79} P.G. de Gennes, {\it Scaling Concepts in Polymer
Physics}, Cornell University Press (1979)

\bibitem{deGe75} P.G. de Gennes, J. Physique Lettres, {\bf 36}, L55 (1975)

\bibitem{Priv86} V. Privman, J. Phys. {\bf A18}, 3281 (1985)

\bibitem{Sale86} H. Saleur, J. Stat. Phys. {\bf 45}, 419 (1986)

\bibitem{Coni86} A. Coniglio, N. Jan, I. Maijd and H.E. Stanley, Phys. Rev.
{\bf B35}, 3617 (1987)

\bibitem{Seno88} F.Seno and A.L. Stella, J. Physique {\bf 49}, 739 (1988)

\bibitem{Madr90} N. Madras, C.E. Soteros, S.G. Whittington, J.L. Martin, 
M.F. Sykes, S. Flesia and D.S. Gaunt, J. Phys. {\bf A23}, 5327 (1990)

\bibitem{Fles92} S. Flesia, D.S. Gaunt, C.E. Soteros and S.G. Whittington,
J. Phys. {\bf A25}, L1169 (1992)

\bibitem{Van93} C. Vanderzande, Phys. Rev. Lett. {\bf 70}, 3595 (1993)

\bibitem{Seno94} F. Seno and C. Vanderzande, J. Phys. {\bf A27}, 5813 (1994);
Corr. {\bf A27}, 7937 (1994)

\bibitem{Stel94} A.L. Stella, Phys. Rev. {\bf E50}, 3259 (1994)

\bibitem{Fles94} S. Flesia, D.S. Gaunt, C.E. Soteros and S.G. Whittington,
J. Phys. {\bf A27}, 5831 (1994)

\bibitem{Moore77} M.A. Moore, J. Phys. {\bf A10}, 305 (1977)

\bibitem{Volk77} M.V. Volkstein, {\it Molecular Biophysics}, Academic 
(New York, 1977)

\bibitem{Obu86} S.P. Obukhov, J. Phys. {\bf A9}, 3655 (1986)

\bibitem{Cardy90} J.L. Cardy, ``Conformal Invariance and Statistical
Mechanics'' in {\it Fields, Strings and Critical Phenomena},
Les Houches XLIX, E. Br\'ezin and J. Zinn-Justin Eds,
North Holland Amsterdam (1990) 

\bibitem{Christe93} P. Christe and M. Henkel, {\it Introduction to Conformal
Invariance and its Applications to Critical Phenomena}, Springer Verlag
Heidelberg (1993)

\bibitem{Nienhuis87} B. Nienhuis, ``Coulomb gas formulation of two
dimensional phase transitions'' in {\it Phase transitions and Critical
Phenomena} vol 11, C. Domb and J.L. Lebowitz Eds, Academic Press (1987)

\bibitem{DS87} B. Duplantier and H. Saleur, Phys. Rev. Lett. {\bf 59},
539 (1987)

\bibitem{VSS91} C. Vanderzande, A.L. Stella and F. Seno, Phys. Rev.
Lett. {\bf 67}, 2757 (1991)

\bibitem{PS81} G. Parisi and N. Sourlas, Phys. Rev. Lett. {\bf 46}, 
871 (1981)

\bibitem{Miller93} J. Miller and K. De'Bell, J. Physique {\bf I 3}, 
1717 (1993)

\bibitem{Giri77} M. Giri, M.J. Stephen and G.S. Grest,
Phys. Rev. {B16}, 4971 (1977)

\bibitem{Wu82} F.Y. Wu, J. Stat. Phys. {\bf 18}, 115 (1978); 
                        Rev. Mod. Phys. {\bf 54}, 235 (1982)

\bibitem{Coni83} A. Coniglio, J. Phys. {\bf A16}, L187 (1983)

\bibitem{Dick84} R. Dickman and W.C. Shieve, J. Physique {\bf 45}, 
1727 (1984)

\bibitem{Lam87} P.M. Lam, Phys. Rev. {\bf B36}, 6988 (1987)

\bibitem{Lam88} P.M. Lam, Phys. Rev.  {\bf B38}, 2813 (1988)

\bibitem{Chang88} I.S. Chang and Y. Shapir, Phys. Rev. {\bf B38}, 
6736 (1988)

\bibitem{Derr83} B. Derrida and H.J. Herrmann, J. Physique {\bf 44}, 1365 
(1983)

\bibitem{Derr82} B. Derrida and L. De Seze, J. Physique {\bf 43},
475 (1982)

\bibitem{Pear95} P. Peard, PhD Thesis, King's College, London (1995)

\bibitem{Cardy95} D. Bennett-Wood, J.L. Cardy, S. Flesia, 
A.J. Guttmann and A. L. Owczarek, to appear J. Phys. {\bf A}, (1995)

\bibitem{Annaka82} A. Annaka and T. Tanaka, Nature {\bf 355}, 430
(1982)

\bibitem{Henk95} M. Henkel, F. Seno and J.M. Yeomans, Oxford preprint 
OUTP-95-56S, submitted to Europhys. Lett. 

\bibitem{HL81} A.B. Harris and T.C. Lubensky, Phys. Rev. {\bf B24},
2656 (1981)

\bibitem{Baxt82} R.J. Baxter, {\it Exactly Solved Models in Statistical
Mechanics}, Academic Press (London 1982), Chap. 4

\bibitem{Gujr95} P.D. Gujrati, Phys. Rev. Lett. {\bf 74}, 1367 (1995); 
J. Chem. Phys.{\bf 98}, 1613 (1993)

\bibitem{Bur82} T.W. Burkhardt and J.M.J. van Leeuwen, 
``Progress and Problems
in Real-Space Renormalization'' in {\it Real Space Renormalization},
T.W. Burkhardt and J.M.J. van Leeuwen Editors, 
Springer-Verlag (1982), p. 1; \\
T.W. Burkhardt, ``Bond-Moving and Variational Methods in Real-Space 
Renormalization'', {\it ibid.}, p. 33

\bibitem{Kauf81} M. Kaufman, R.B. Griffiths, J.M. Yeomans and M.E. Fisher,
Phys. Rev. {\bf B23}, 3448 (1981)

\bibitem{Bara83} A. Baracca, M. Bellesi, R. Livi, R. Rechtman and
S. Ruffo, Phys. Lett. {\bf 99A}, 156 (1983)

\bibitem{Itz89} C. Itzykson and J.-M. Drouffe, {\it Statistical Field
Theory}, Vol. 1, Cambridge University Press (1989), Chap. 4

\bibitem{Dhar87} D. Dhar, J. Phys. {\bf A20}, L847 (1987)


\end{references}
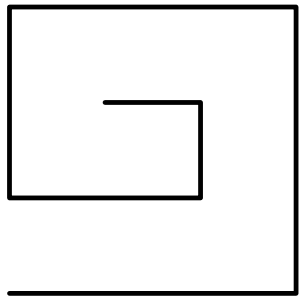
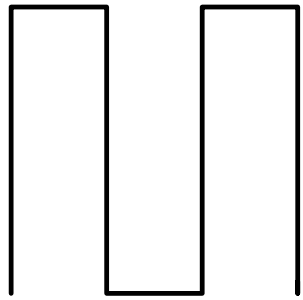
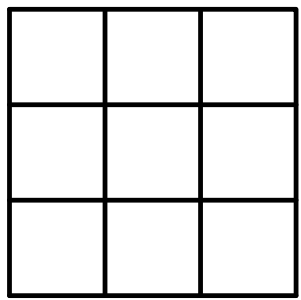
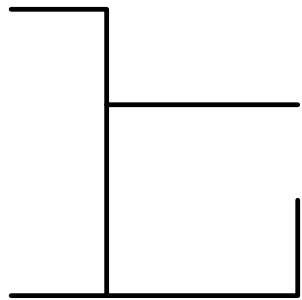
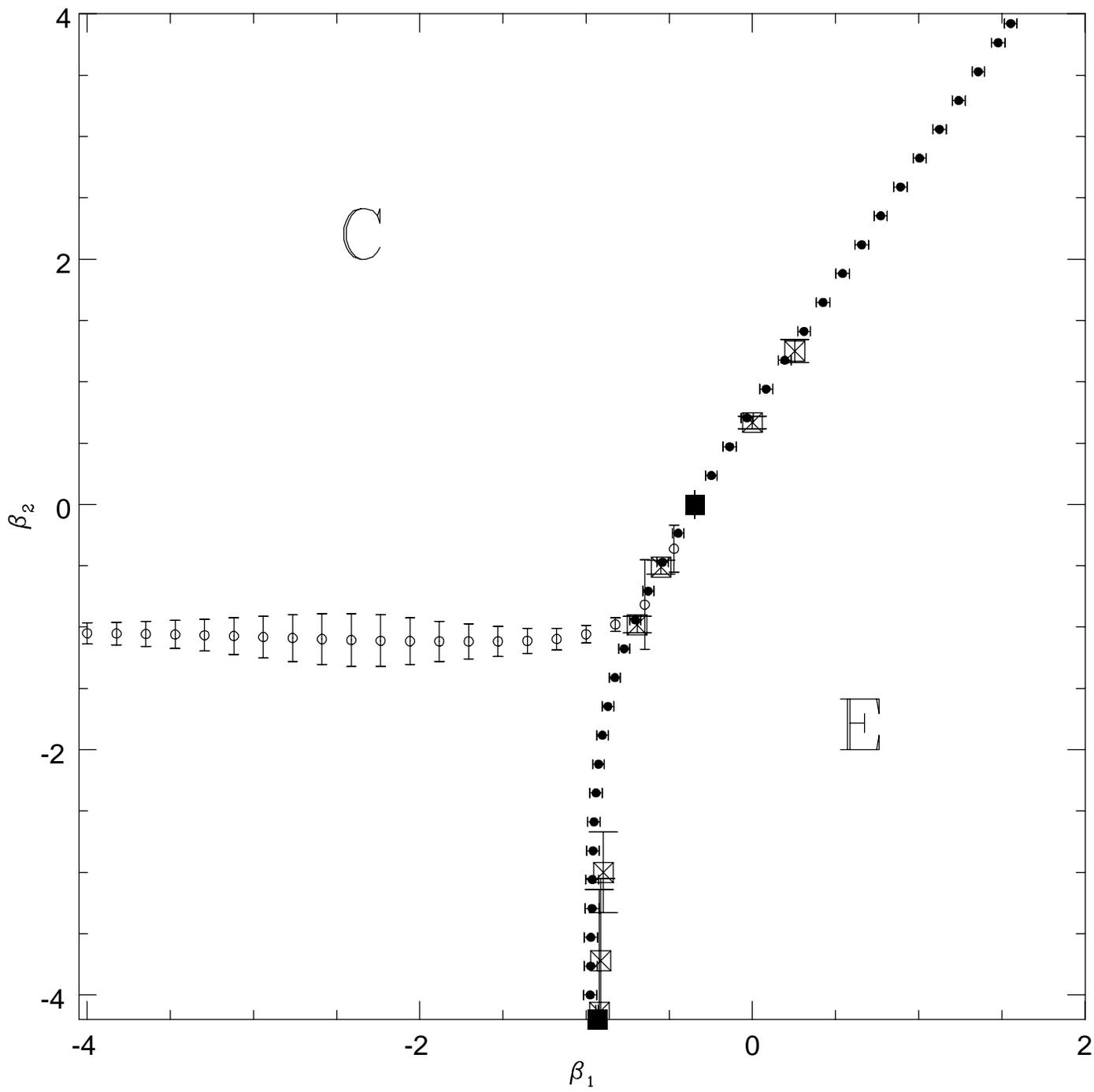
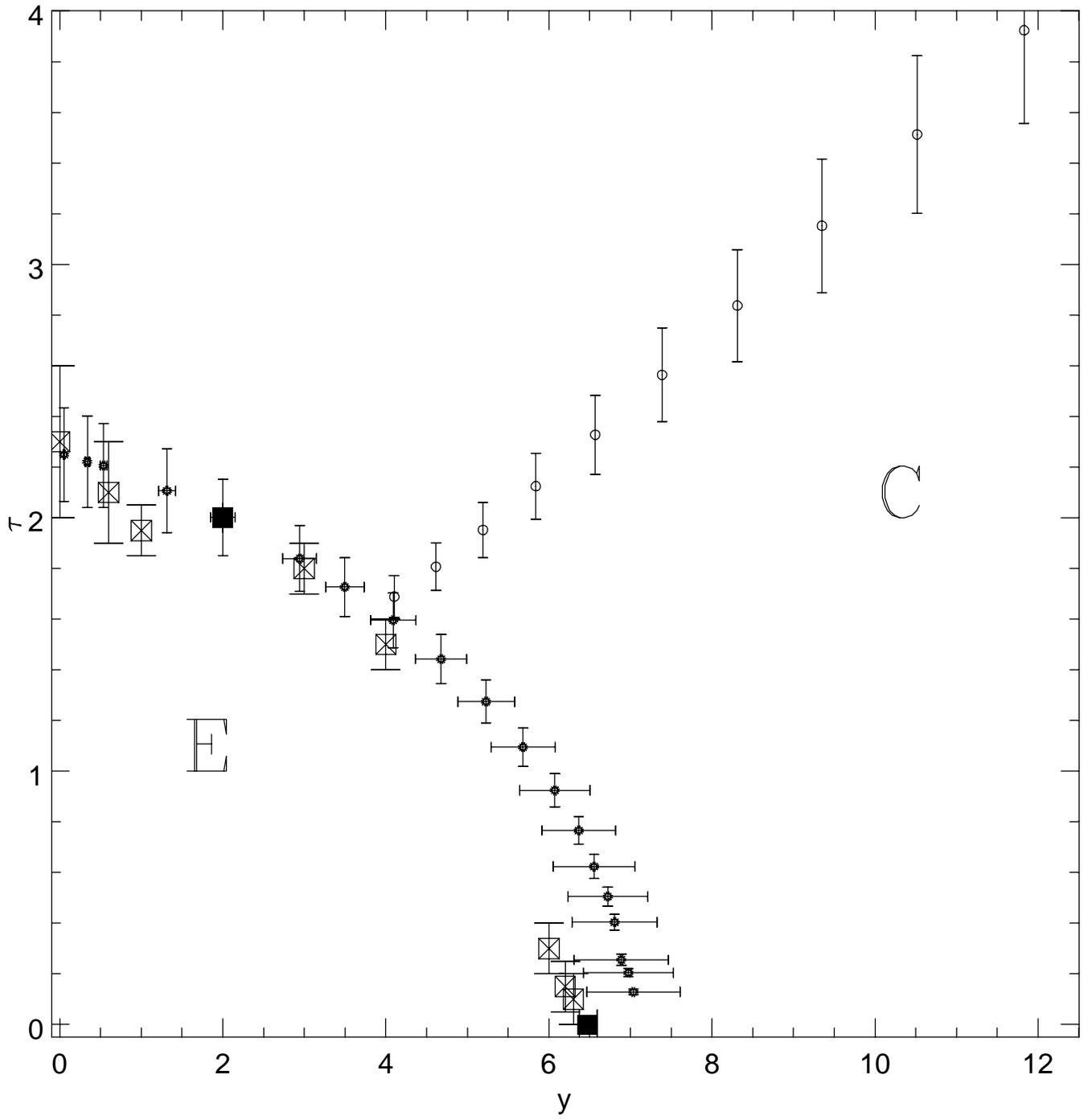
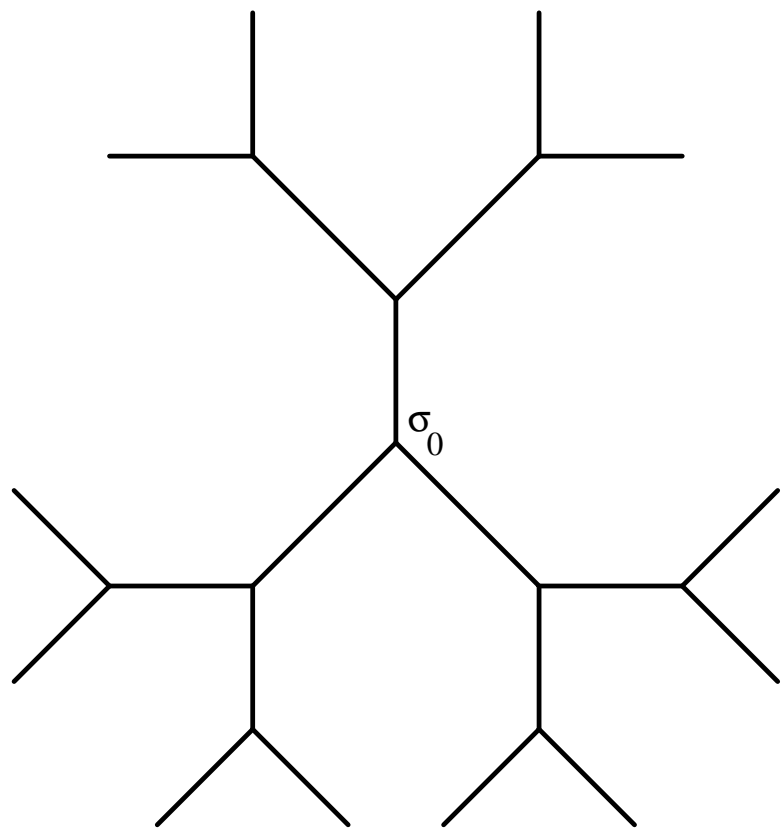
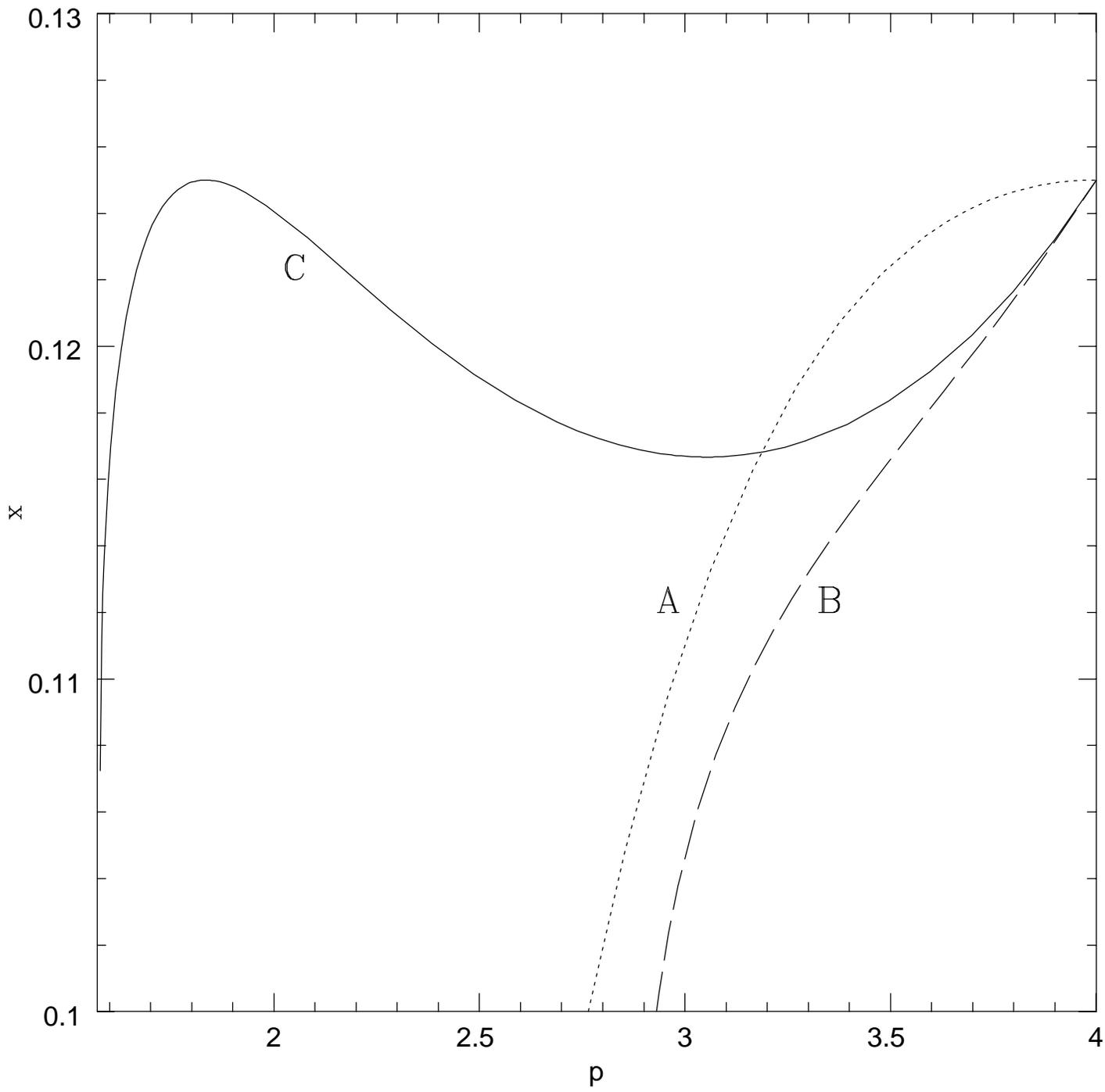
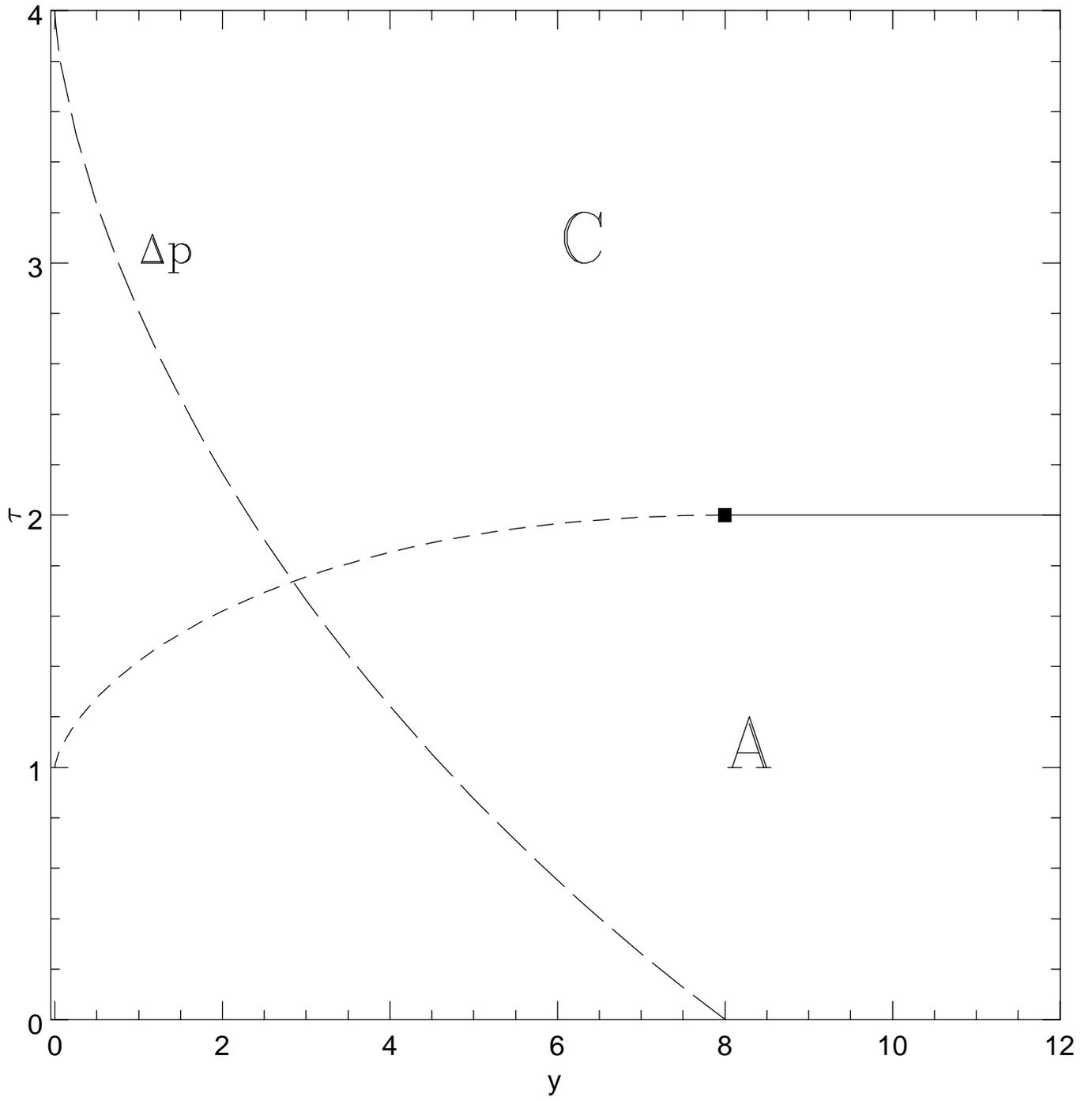
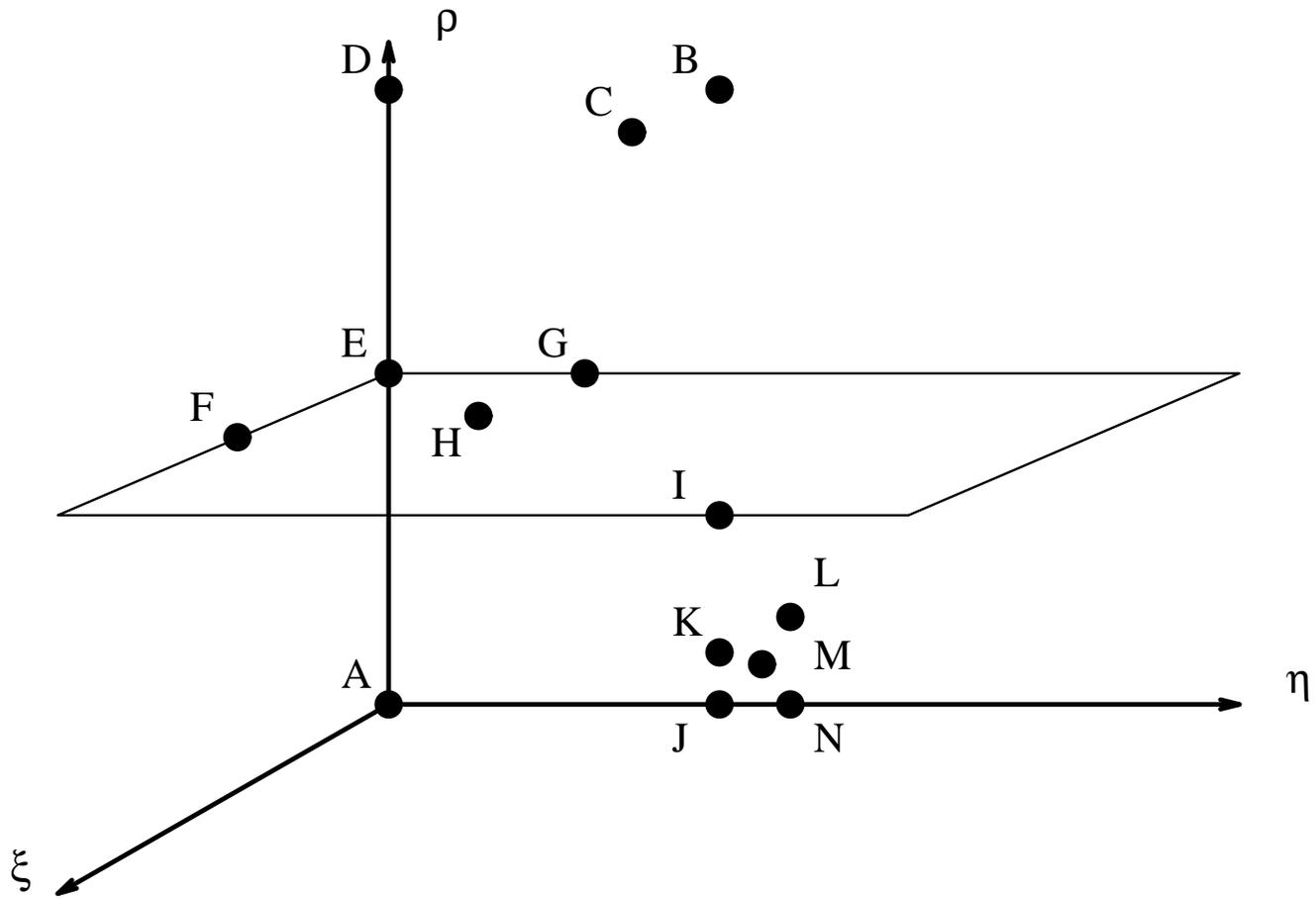
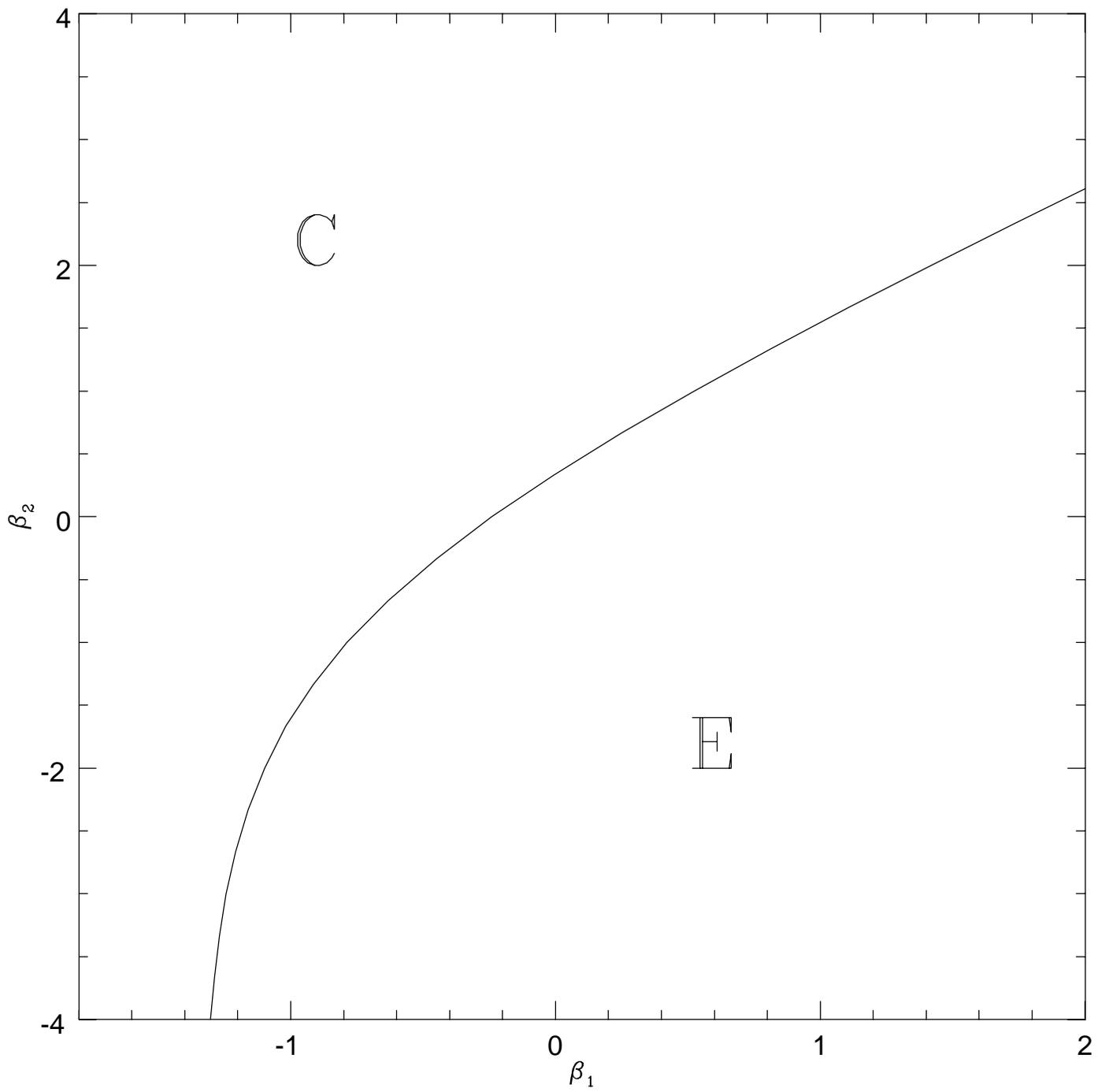
\end{document}